\pdfoutput=1
\documentclass[a4paper,12pt]{article}
\usepackage[colorlinks=true, urlcolor=blue]{hyperref}
\usepackage[usenames,dvipsnames]{color}
\input xy
\xyoption{all}
\hypersetup{
    bookmarks=true,         
    unicode=false,          
    pdftoolbar=true,        
    pdfmenubar=true,        
    pdffitwindow=true,      
    pdftitle={Fully Dynamic Algorithm for Maximum matching},    
    pdfauthor={Ankit Sharma, Manoj Gupta, Surender Baswana},     
    pdfsubject={Subject},   
    pdfcreator={Creator},   
    pdfproducer={Producer}, 
    pdfkeywords={keywords}, 
    pdfnewwindow=true,      
    colorlinks=true,       
    linkcolor=blue,          
    citecolor=green,        
    filecolor=magenta,      
    urlcolor=blue           
}
\usepackage{color}
\definecolor{light-gray}{gray}{0.85}
\usepackage{fancyhdr}
\usepackage{colortbl}
\usepackage[cm]{fullpage}
\title{An O(log(n)) Fully Dynamic Algorithm for Maximum matching in a tree}
\author{Manoj Gupta\footnote{gmanoj@iitk.ac.in}, Ankit Sharma\footnote{ankitsh@iitk.ac.in}\\Indian Institute of Technology Kanpur, India}
\begin{document}
\maketitle
\begin{abstract}
In this paper, we have developed a fully-dynamic algorithm for maintaining cardinality of maximum-matching in a tree using the construction of top-trees. The time complexities are as follows:
\begin{enumerate}
\item Initialization Time: $O(n(log(n)))$ to build the Top-tree.
\item Update Time: $O(log(n))$
\item Query Time: $O(1)$ to query the cardinality of maximum-matching and $O(log(n))$ to find if a particular edge is matched.
\end{enumerate}
\end{abstract}
\section{Introduction}
Dynamic graph algorithms aim to maintain certain properties in a graph under insertion and/or deletion of edges or vertices from the graph. The motivation behind dynamic algorithms is to maintain the graph property without the need to compute from scratch after each update (insertion or deletion). A dynamic algorithm is incremental and decremental if it can handle the cases of insertion and deletion respectively. A dynamic algorithm is fully-dynamic if it is both incremental and decremental. Dynamic graph algorithms for maintaining connectivity and mimimum spanning tree in a graph have been well studied.

This paper presents a dynamic algorithm for maintaining maximum matching in a dynamic tree. Given a graph $G (V,E)$, maximum matching gives the subset of $E$ of maximum cardinality such that no two edges in the subset share a vertex. The best static algorithm for finding maximum matching in a general graph has $O(|E|\sqrt{|V|})$ time complexity \cite{maxmatch-static}. The static algorithm for finding maximum-matching in a tree has $O(n)$ time complexity. The algorithm involves randomly choosing a leaf node and making the edge, incident on the leaf, matched. The matched edge is then deleted from the tree along with all the edges which are incident on the node to which the leaf node was attached. The process is recursively followed on the new tree till no nodes are left. The time complexity of the algorithm is $O(n)$, $n$ being the number of nodes in the tree, since in each iteration atleast one node is removed and number of edges in a tree are $n-1$. We omit the proof of correctness of this algorithm.

Among dynamic algorithm for maximum-matching in a graph, the best time complexity is of a randomized algorithm given by Piotr Sankowski which has time complexity of $O(n^{1.495})$ \cite{maxmatch-dynamic}. In this paper, we give an $O(\log n)$ time algorithm to maintain maximum matching in a dynamic tree under insertion and deletion of edges, $n$ being the number of nodes in the tree. We make use of top-tree, introduced in section \ref{sec:top-tree}, to represent the dynamic tree. The algorithm, the proof of correctness and time-complexity analysis are presented in subsequent sections.

\section{\label{sec:top-tree}Top Tree}
Top-tree\cite{werneck} partitions a tree into clusters, a cluster being defined as a connected subtree of the given tree. In figure \ref{fig:partition-example}, $P$, $Q$ and $R$ are clusters of the tree. A top-tree has several levels and each level induces a partitioning of the tree into clusters. A higher level in a top-tree partitions the tree into smaller number of larger clusters compared to a lower level. To construct level $l+1$ from $l$, the clusters at level $l$ are combined according to prescribed rules to form the clusters at the next level in the top tree. At the lowest level, each edge of the tree is a cluster. At the highest level, the whole tree is seen as a single cluster. The number of levels in a top-tree is of $O(log(n))$ where $n$ denotes the number of nodes in the tree. The reason behind $O(log(n))$ levels is that the cluster combination operations and the fact that we are dealing with a tree ensure that each successive level has atmost a constant ratio of number of clusters of the previous level.

\begin{figure}
\begin{tabular}{lllll}
\xymatrix{
&&&&g\ar@{.}@/^/[dd] \ar@{.}@/_/[llld]&&J \ar@{.}@/_1pc/[lldd] \ar@{.}@/^1pc/[lldd]|{}="r"\\
&b\ar@{-}[dr] \ar@{.}@/_/[dddr]&&f\ar@{-}[ur] \ar@{-}[dr]&&i\ar@{-}[ur]\\
a\ar@{-}[ur] \ar@{.}@/_/[ur]|{}="p"  \ar@{.}@/^/[ur]&&c\ar@{-}[ur] \ar@{-}[dr]&&h\ar@{-}[ur]&&R\ar@{->}"r"\\
&P\ar@{->};"p"&&d\ar@{-}[dl]\\
&&e\ar@{.}@/_1pc/[rruu]|{}="q"&&Q\ar@{->}"q"&&\\
}&&&&
\xymatrix{
&b\ar@{-}[dr]&&j\\
a\ar@{-}[ur]&&h\ar@{-}[ur]&&\\
}
\end{tabular}
\caption{The figure on the left shows partition of the tree into three clusters $P$, $Q$ and $R$, having boundary nodes $(a,b)$, $(b,h)$ and $(h,j)$ respectively. For instance, cluster $Q$ is the subtree spanned between $b$ and $h$. Clusters $P$ and $R$ are leaf-clusters. The figure on the right shows the equivalent top-tree representation of the tree partition wherein each cluster of the tree is denoted by an edge with end points of the edge being the boundary nodes of the corresponding cluster. For example, cluster $Q$ is represented by an edge having end points $b$ and $h$.}
\label{fig:partition-example}
\end{figure}

The cluster formation rules ensure that each cluster is connected to the rest of the tree at atmost two nodes. These two nodes are referred to as `boundary nodes' of the given cluster. Those clusters which share only one node with the rest of the tree are called `leaf-clusters'. Although a leaf cluster has only one node shared with the rest of the tree and therefore in principle should have only one boundary node, yet for the sake of uniformity across all clusters, we assign two nodes of a leaf-cluster as boundary nodes. One of the boundary nodes is the one through which the leaf-cluster is connected to the rest of the tree. The way the other boundary node is decided will be clear from discussion in the following paragraphs. At the lowest level of the top-tree, each edge is a cluster. The leaf edges of the tree are leaf-clusters. The boundary nodes for all clusters are the end-points of the edge forming the cluster.

As mentioned earlier, clusters, at level $l$, are combined to form clusters at level $l+1$ in a top-tree. Each cluster at level $l+1$ is either a combination of two clusters of level $l$ or is same as a cluster of level $l$. To form level $l+1$ clusters, we combine clusters of level $l$ in pairs to form level $l+1$ clusters till we cannot combine any more clusters. Each cluster of level $l$ can participate in atmost one combine operation with another cluster of the same level to form a cluster at level $l+1$.

At any level, more than one cluster may share a common boundary node. In this case, we have an ordering of the clusters around the boundary node which is in counterclockwise orientation around the node. In the example below, clusters $P$, $Q$ and $R$ share the boundary node $B$; $P$ is the successor of $R$ and predecessor of $Q$. We call a cluster `incident' on a node if the node is a boundary node of the cluster and the node is a boundary node of atleast one more cluster.

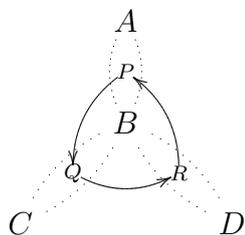
\xymatrix{
&A\\
&B \ar@{}[u]|{P}^{}="u" \ar@{.}@/_/[u] \ar@{.}@/^/[u] \ar@{}[ld]|{Q}^{}="l" \ar@{}[rd]|{R}^{}="r" \ar@/_/"u";"l" \ar@/_/"l";"r" \ar@/_/ "r";"u"   \ar@{.}@/_/[ld] \ar@{.}@/^/[ld] \ar@{.}@/_/[rd] \ar@{.}@/^/[rd]&\\
C&&D
}
Two clusters can be combined only if they have a successor-predecessor relationship. Further, two clusters are combined either by a `rake' operation or a `compress' operation.
\begin{figure}
\begin{tabular}{ccc}
\xymatrix{
&B \ar@{.}@/_/[ld] \ar@{..}@/^/[ld] \ar@{.}@/_/[rd] \ar@{.}@/^/[rd] \ar@{}[ld]|{P}="p" \ar@{}[rd]|Q="q"\\
A&&C \ar@/_/"p";"q"\\
}&
\xymatrix{
\ar@{=>}[rr]^{\textrm{rake}}&&
}&
\xymatrix{
&B \ar@{.}@/^/[rd] \ar@{.}@/_3pc/[rd] \ar@{}[rd]|{R}\\
&&C
}
\end{tabular}
\caption{Example of Rake operation: Here cluster $P$ is raked onto cluster $Q$ to form cluster $R$.}
\label{fig:rake-example}
\end{figure}
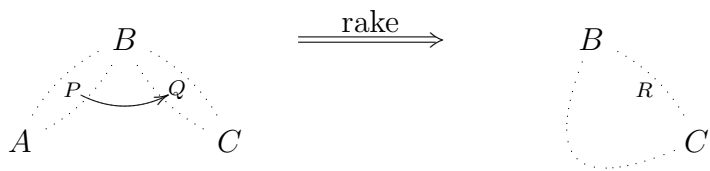
\begin{figure}
\begin{tabular}{ccc}
\xymatrix{
A \ar@{}[rr]|{P}="p" \ar @/^1pc/ @{.} [rr] \ar@/_1pc/@{.}[rr]& & B \ar@{}[rrr]|{Q}="q" \ar@{<->}@/_1pc/"p";"q"\ar @/^1pc/ @{.} [rrr] \ar@/_1pc/@{.}[rrr]& & & C\\
}&
\xymatrix{
\ar@{=>}[rr]^{\textrm{compress}}&&
}&
\xymatrix{
A\ar@{}[rrr]|{R} \ar @/^1pc/ @{.} [rrr] \ar@/_1pc/@{.}[rrr]&  & & C\\
}
\end{tabular}
\caption{Example of compress operation: Here clusters $P$ and $Q$ are compressed to form cluster $R$.}
\label{fig:compress-example}
\end{figure}
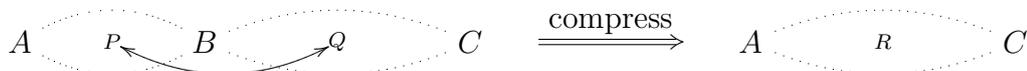

\begin{enumerate}
\item Rake operation: One of the two clusters is necessarily a leaf cluster and the leaf cluster should be the predecessor of the other cluster. As shown in diagram, cluster $P$ has boundary nodes $A$ and $B$ and cluster $Q$ has boundary nodes $B$ and $C$. Cluster $P$ is a leaf cluster and is connected to the rest of the tree through $B$. Cluster $Q$ may or may not be a leaf cluster and is connected to rest of the tree through $B$ and $C$. Cluster $P$ is `raked onto' cluster $Q$ to form cluster $R$ with boundary nodes $B$ and $C$. Cluster $R$ contains all the edges and nodes which are part of clusters $P$ and $Q$.\\

\item Compress operation: Here the two clusters, sharing a common boundary node, should be the only clusters incident on the shared boundary node. As shown in diagram, cluster $P$ has boundary nodes $A$ and $B$ and cluster $Q$ has boundary nodes $B$ and $C$. Further, the shared boundary node $B$ has no cluster incident on it other than $P$ and $Q$. Clusters $P$ and $Q$ are `compressed' to form cluster $R$ with boundary nodes $A$ and $C$. Cluster $R$ contains all the edges and nodes which are part of clusters $P$ and $Q$.\\
\end{enumerate}

\begin{figure}
\begin{tabular}{ll}
\xymatrix{
&l=0 &&& l=1 &&& l=2 \\
a \ar@{-}[dr] ^{}="a1" && b \ar@{-}[dl] ^{}="b1" \ar@/_/ "b1";"a1"|{r} && a \ar@{-}[d] ^{}="a2" &&& a \ar@{-}[d]  \\
& c \ar@{-}[d]^{}="c1" &&& c \ar@{-}[d]^{}="c2" \ar@{<->}@/_/|{c} "c2";"a2" &&& e\\
& d \ar@{-}[d]^{}="d1" &&& e\\
& e \ar@{<->}@/_/|{c} "d1";"c1"}&Top tree\\
\xymatrix{
\\
a\ar@{-}[dr] \ar@{.}@/_/[dr] \ar@{.}@/^/[dr] &&b\ar@{-}[dl] \ar@{.}@/_/[dl] \ar@{.}@/^/[dl] &&a\ar@{-}[dr] \ar@{.}@/_/[dr] &&b\ar@{-}[dl] \ar@{.}@/_/[ll] &a\ar@{-}[dr] \ar@{.}@/_/[dddr] &&b\ar@{-}[dl] \ar@{.}@/_/[ll] \\
&c\ar@{-}[d] \ar@{.}@/_/[d] \ar@{.}@/^/[d] &&&&c\ar@{-}[d] \ar@{.}@/_/[ur]  \ar@{.}@/_/[dd] \ar@{.}@/^/[dd] &&&c\ar@{-}[d]\\
&d\ar@{-}[d] \ar@{.}@/_/[d] \ar@{.}@/^/[d] &&&&d\ar@{-}[d]&&&d\ar@{-}[d]\\
&e&&&&e&&&e\ar@{.}@/_/[uuur] 
}&\begin{tabular}{l}Underlying partitioning\\of the tree into clusters\end{tabular}
\end{tabular}
\caption{This figure shows top-tree for the tree given at l=0 level. In the upper half of the figure, the different levels of the top-tree are shown. The lower half shows the underlying partition of the tree into clusters at each level. The arrows shown in the construction of top-tree are labeled by $r$ and $c$ for rake and compress operation respectively.}
\label{top-tree-example}
\end{figure}
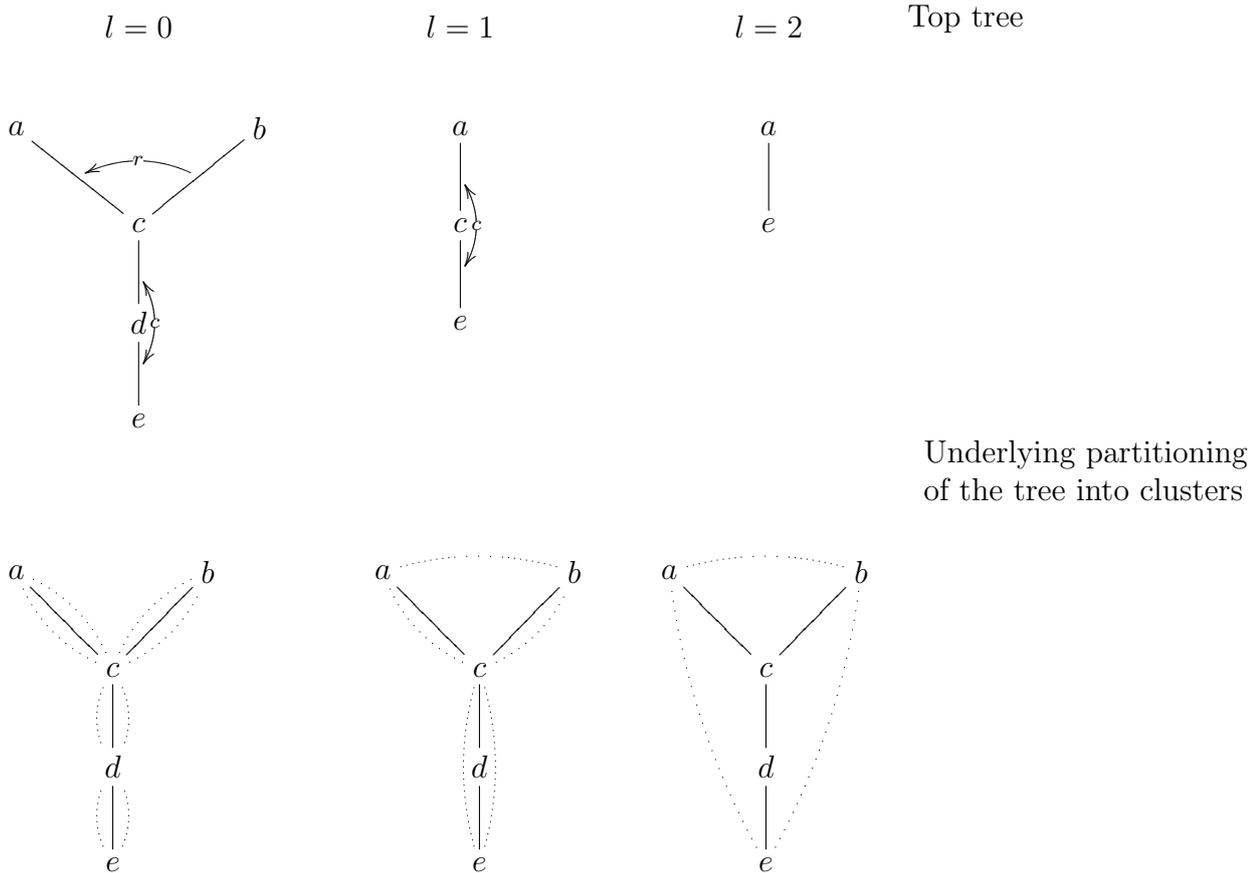

Top-tree represents each cluster, at a given level, by an edge with end-points of the edge being the boundary nodes of the cluster. For the clusters which share a boundary node, the edges, corresponding to the clusters, likewise share the common boundary node. Hence, at each level the cluster partition is represented by a tree or a forest formed of the edges representing the clusters. The top tree for a tree is shown in figure \ref{top-tree-example}. The tree is not shown explicitly in the figure and is same as the tree shown at level 0 of the top-tree. Let $(x,y)_i$ denote the edge at level $i$. At level 0, we have the original tree. The edge $(b,c)_{0}$ is raked onto $(a,c)_{0}$ around $c$ to give $(a,c)_{1}$. The edge $(c,d)_{0}$ and $(d,e)_{0}$ is compressed at $d$ to give $(c,e)_{1}$. These are the only operations at this level. Hence, $(a,c)_{1}$ is the parent of $(a,c)_0$ and $(b,c)_0$ and $(c,e)_1$ is the parent of $(c,d)_0$ and $(d,e)_0$. Cluster-combination operations are performed recursively to give successive levels and these levels form the top-tree for the given tree. 

\subsection{Updation in a Top Tree}
Updation in a top-tree under insertion or deletion of edges can be performed in $O(log(n))$ time. Addition or deletion of an edge makes some of the cluster-combination operations performed in the original top-tree invalid in the modified tree. Further, certain new combination operations might arise. We undo the invalid operations of the original top-tree and reconstruct the new top-tree level-by-level.

\begin{figure}
\xymatrix{
&l=0 &&& l=1 && l=2 && l=3\\
a\ar@{-}[dr] ^{}="a1" && b \ar@{-}[dl] ^{}="b1"  \ar@/_/|{r} "b1";"a1"   && a\ar@{-}[d] ^{}="a2" && a\ar@{-}[d] ^{}="a3" && a\ar@{-}[d]\\
& c \ar@{-}[d]^{}="c1" &&& c \ar@{-}[d] ^{}="c2"   \ar@{<->}@/_/|{c} "c2";"a2"  && d \ar@{-}[d] ^{}="d3" && f\\
& d \ar@{-}[dl]^{}="d1" \ar@{-}[dr]^{}="e1" &&& d \ar@{-}[d] ^{}="d2" && f \ar@{<->}@/_/|{c} "d3";"a3" \\
e && f \ar@/_/|{r}"d1";"e1"  && f}
\caption{This figure shows the construction of the new top-tree on addition of the edge $(d,f)$. The arrows shown in the construction of top-tree are labeled by $r$ and $c$ for rake and compress operation respectively.}
\label{fig:top-tree-updation}
\end{figure}
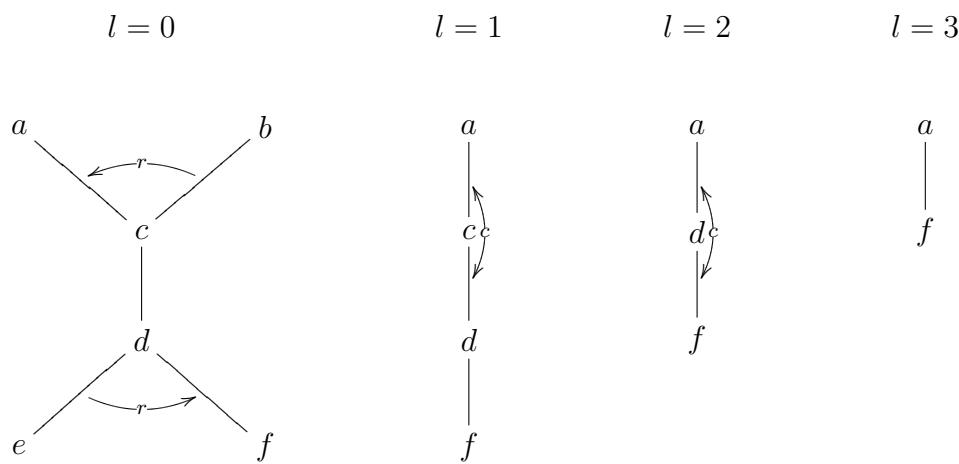

We explain the updation operation in a top-tree with the help of an example. For a more detailed description and the proof for its time complexity, we refer the reader to the Renato F. Werneck's thesis\cite{werneck}. Let us take the tree constructed in the earlier section and look at what happens when an edge $(d,f)$ is added at the lowest level. The new top tree is given in Figure \ref{fig:top-tree-updation}. Let us see how we can transform the original top tree above to top tree below.

We have to add cluster $(d,f)_0$ at the lowest level. This makes the compress operation of $(c,d)_{0}$ and $(d,e)_{0}$, made in the original top-tree, invalid as the common shared boundary node $d$ now has and additional cluster $(d,f)_{0}$ incident on it. As the original compress operation is invalid, hence at level $1$, $(c,e)_1$ is deleted. Further, a new rake operation of $(e,d)_{0}$ onto $(d,f)_{0}$ is performed to give $(d,f)_{1}$ at level $1$. Edge $(c,d)_1$ is also added at level $1$ as $(c,d)_{0}$ could not participate in any combine operation at level $0$. At level $2$, edge $(a,e)_2$ is deleted since its child $(c,e)_1$ has been deleted. Also, edges $(a,d)_2$ and $(d,f)_2$ are added. Similarly, we build rest of the top-tree.

The construction of the top tree ensures that there are only a constant number of addition and deletion operations at each level. Hence, the time to update the top tree is $O(\log n)$. The intuitive reason behind the constant number of operations at each level is that each cluster is connected to the rest of the tree at atmost 2 points which means that any change to this cluster would affect only a constant number of clusters at that level namely the clusters which are the predecessor or successor of the cluster at its two boundary nodes. This completes our discussion on top-trees.

\section{Algorithm}
\subsection{Observations\label{observations}}
We now present an $O(log(n))$ update time fully dynamic algorithm for maintaing maximum-matching in a tree. We first present a few observations which aid in the development of the algorithm and then the algorithm itself.

{\bf Observation 1:} If we are able to maintain certain information with each cluster in a top-tree such that the information for a parent cluster can be derived from the information stored in its child clusters, then any change in information due to insertion or deletion of an edge at the bottom-most level of the top-tree can be propagated up the top-tree in as much time as the update operation takes for a top-tree. This observation motivates us to store some information with each cluster such that it follows the property mentioned above and that it helps in maintaining the maximum-matching.

Before we present the information which we store with each cluster, let us introduce a few notations to aid us in this task. In a tree $T$, let $T_{E}$ denote the set of edges contained in $T$. We call a node `matched' if any of the edges incident on it are matched. If none of the edges incident on a node is matched, we call the node `unmatched'. If $Q$ be a subtree of tree $T$ and $M$, a particular matching of $T$, let $A_{Q,M}$ denote the status - matched or unmatched - of node $A$ under matching $M$ when restricted to the subtree $Q$ i.e. we consider only those edges which are in $Q_{E}$ and are incident on $A$ and in case, any of these edges under consideration are matched under M, we put $A_{Q,M}$ as matched and unmatched otherwise.


In the diagram below, a particular matching $M$ is shown for tree $T$ wherein the dashed edges are matched and the solid edges are unmatched. $Q$ is the subtree spanning between $A$ and $B$. $A_{Q,M}$ is unmatched since the only edge under consideration A--q1 is unmatched. Further, $A_{T,M}$ is matched since out of the two edges, A---t1 and A---q1, under consideration, A---t1 is matched. Again, $B_{Q,M}$ is matched since B---q3, the only edge under consideration, is matched and $B_{T,M}$ is also matched since out of the two edges B---q3 and B---t2, under consideration, B---q3 is matched.
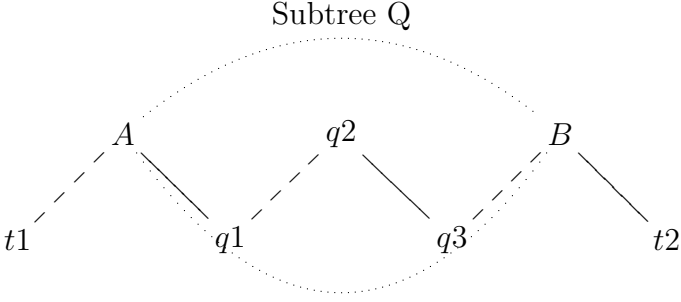
\begin{tabular}{c}
\xymatrix{
\\
& A\ar @{-}[dr] \ar @{.}@/^3pc/ [rrrr]^{\textrm{Subtree Q}} \ar @{.}@/_5pc/ [rrrr]& & q2 \ar @{-}[dr]& & B\ar @{-}[dr] & \\
t1\ar @{--}[ur]& & q1 \ar @{--}[ur]& & q3 \ar @{--}[ur] & &t2 \\
\\
}
\end{tabular}

If a node is matched, we call it black ($b$) and if it is unmatched, we call it white ($w$). In the previous example, $A_{Q,M}=w$, $A_{T,M}=b$, $B_{Q,M}=b$ and $B_{T,M}=b$.

As mentioned earlier, each cluster in a top-tree is a sub-tree of the tree $T$ such that it shares atmost two nodes with the rest of the tree, which we call the boundary nodes. Let us take a particular cluster, say $Q$, having boundary nodes $A$ and $B$. Let us consider the set of matchings in $Q$, which make node $A$ matched ($b$) and $B$ unmatched ($w$) and consider the matching $N$ which has the maximum cardinality in this set. We denote the cardinality of this particular matching $N$ by $M_{Q}^{b_{A}w_{B}}$ where $b_{A}$ and $w_{B}$ denote that $A$ is matched ($b$) and $B$ is unmatched ($w$). $M_{Q}^{w_{A}w_{B}}$, $M_{Q}^{w_{A}b_{B}}$ and $M_{Q}^{b_{A}b_{B}}$ are similarly defined.

We now return to our orginal question of what information should be maintained with each cluster in a top-tree such that it helps in maintaining maximum-matching and follows the property that the information of the parent can be deduced from that of its children. With each cluster $Q$, we maintain the following information: $M_{Q}^{b_{A}w_{B}}$, $M_{Q}^{w_{A}w_{B}}$, $M_{Q}^{w_{A}b_{B}}$ and $M_{Q}^{b_{A}b_{B}}$. 

The reason behind this choice of information is presented below.  Essentially, this information fulfills the two critera mentioned above. First, the information is sufficient to calculate maximum-matching as the cardinality of maximum-matching of $Q$, denoted by $M_{Q}$, is the maximum number among $M_{Q}^{b_{A}w_{B}}$, $M_{Q}^{w_{A}w_{B}}$, $M_{Q}^{w_{A}b_{B}}$ and $M_{Q}^{b_{A}b_{B}}$ since these four cases exhaust the set of possibilities of matchings cases of $A$ and $B$.

Secondly, this information obeys the property that the information of the parent cluster can be calculated using the information stored in the child-clusters. To see how, we present the following observation.

{\bf Observation 2:} Let us consider two trees $P$ and $Q$ sharing a common node $B$ as shown in figure. The internal structure of the subtrees is not shown in the figure. 

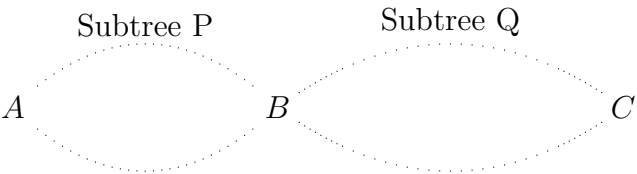
\xymatrix{
\\
\\
A\ar @/^2pc/ @{.} [rrr]^{\textrm{Subtree P}} \ar@/_2pc/@{.}[rrr]& & & B \ar @/^2pc/ @{.} [rrrr]^{\textrm{Subtree Q}} \ar@/_2pc/@{..}[rrrr]& & & & C
\\
\\
}

Let tree $T$ denote the union of subtrees $P$ and $Q$. Say we take a matching $M$ of $P$ and a matching $N$ of $Q$ and produce a matching $O$ of $T$ as follows -- under matching $O$, an edge $e \in T_{E}$ follows matching imposed by $M$ if $e \in P_{E}$ and follows matching imposed by $N$ if $e \in Q_{E}$. Clearly, we cannot combine any two arbitrary matchings $M$ and $N$, since then we can have two matched edges incident on node $B$, one in $Q$ and the other in $P$. Hence, we can combine only those matchings of $P$ and $Q$ which lead to atmost one matched edge incident on $B$, in other words, atmost one of $B_{P,M}$ and $B_{Q,N}$ is black.

Let us suppose we have $M_{P}^{b_{A}w_{B}}$, $M_{P}^{w_{A}w_{B}}$, $M_{P}^{w_{A}b_{B}}$ and $M_{P}^{b_{A}b_{B}}$ for $P$ and $M_{Q}^{b_{B}w_{C}}$, $M_{Q}^{w_{B}w_{C}}$, $M_{Q}^{w_{B}b_{C}}$ and $M_{Q}^{b_{B}b_{C}}$. Can we generate $M_{T}^{w_{A}b_{C}}$ from this available information? We can see that by imposing the constraint that we can combine only those matchings of $P$ and $Q$ which yield atmost one matched edge on $B$ in $T$, we get
\begin{eqnarray}
M_{T}^{w_{A}b_{C}} = max\{M_{P}^{w_{A} b_{B}}+M_{Q}^{ w_{B} b_{C}}, M_{P}^{w_{A} w_{B}}+M_{Q}^{ b_{B} b_{C}}, M_{P}^{w_{A} w_{B}}+M_{Q}^{ w_{B} b_{C}}\}
\end{eqnarray}

In the above equation, we can understand that LHS $\ge$ RHS, as each possibility in RHS is a valid combination of matchings of $P$ and $Q$ inferred from the discussion above. The reason for LHS = RHS is that the options in RHS exhaust the set of possibilities. Similarly, we can obtain $M_{T}^{w_{A}w_{C}}$, $M_{T}^{b_{A}w_{C}}$ and $M_{T}^{b_{A}b_{C}}$. Again, $M_{T}$ would be the maximum of the four numbers.

Hence, we can see that if we know cardinality of certain constrained maximum-matchings for sub-trees $P$ and $Q$, we can generate the cardinality of constrained maximum-matchings for $T$. This fact is used in the algorithm. With each cluster, we maintain the {\em information} of cardinality of the four constrained maximum-matchings. The information of a parent cluster can be calculated using the information of the child-clusters. To obtain the cardinality of maximum-matching for the whole tree, we find the maximum cardinality among the four constrained maximum-matchings for the cluster at the highest level of the top-tree. Further, whenever we add or delete an edge, it is an O(1) time operation to update the information for each modified cluster in the top-tree as the information of the children clusters is enough to calculate information of the parent cluster.

\subsection{Algorithm}
We now describe the algorithm for maintaing maximum-matching in a tree using a top-tree. With each cluster $P$ with boundary nodes, say, $A$ and $B$ we maintain the four values $M_{P}^{b_{A}w_{B}}$, $M_{P}^{w_{A}w_{B}}$, $M_{P}^{w_{A}b_{B}}$ and $M_{P}^{b_{A}b_{B}}$.

For the base case, where each cluster, say $P$, consists of an edge $(A,B)$, we have $M_{P}^{w_{A}w_{B}}$ and $M_{P}^{b_{A}b_{B}}$ are $1$. $M_{P}^{w_{A}b_{B}}$ and $M_{P}^{b_{A}w_{B}}$ are invalid cases since the only edge $(A,B)$ can either be matched or unmatched meaning either both $A$ and $B$ are either matched or unmatched. For the invalid case, we assign a special symbol $null$ i.e. $M_{P}^{b_{A}w_{B}}=null$ and $M_{Q}^{w_{A}b_{B}}=null$. Any addition operation with a $null$ value yields a $null$ value and a maximum operation over a set consisting of $null$ and non-$null$ values should yield maximum value over non-$null$ values if there are any and $null$ otherwise.

Let us now consider how we maintain the values $M_{P}^{u_{A}v_{B}}; u,v \in \{w,b\}$ when we have a rake or a compress operation. We have the following tables. The notation of clusters $P$, $Q$ and $R$ are same as shown in Figures \ref{fig:rake-example} and \ref{fig:compress-example}.

\begin{itemize}
\item Compress
\begin{enumerate}
\item $M_{R}^{ w_{A} w_{C}} = max\{M_{P}^{ w_{A} w_{B}} + M_{Q}^{ w_{B} w_{C}}, M_{P}^{ w_{A} w_{B}} + M_{Q}^{ b_{B} w_{C}}, M_{P}^{ w_{A} b_{B}} + M_{Q}^{ w_{B} w_{C}}\}$
\item $M_{R}^{ w_{A} b_{C}} = max\{M_{P}^{ w_{A} w_{B}} + M_{Q}^{ w_{B} b_{C}}, M_{P}^{ w_{A} w_{B}} + M_{Q}^{ b_{B} b_{C}}, M_{P}^{ w_{A} b_{B}} + M_{Q}^{ w_{B} b_{C}}\}$
\item $M_{R}^{ b_{A} w_{C}} = max\{M_{P}^{ b_{A} w_{B}} + M_{Q}^{ w_{B} w_{C}}, M_{P}^{ b_{A} w_{B}} + M_{Q}^{ b_{B} w_{C}}, M_{P}^{ b_{A} b_{B}} + M_{Q}^{ w_{B} w_{C}}\}$
\item $M_{R}^{ b_{A} b_{C}} = max\{M_{P}^{ b_{A} w_{B}} + M_{Q}^{ w_{B} b_{C}}, M_{P}^{ b_{A} w_{B}} + M_{Q}^{ b_{B} b_{C}}, M_{P}^{ b_{A} b_{B}} + M_{Q}^{ w_{B} b_{C}}\}$
\end{enumerate}

\item Rake
\begin{enumerate}
\item $M_{R}^{ w_{B} w_{C}} = max\{M_{P}^{ w_{A} w_{B}} + M_{P}^{ w_{B} w_{C}}, M_{P}^{ b_{A} w_{B}} + M_{P}^{ w_{B} w_{C}}$\}
\item $M_{R}^{ w_{B} b_{C}} = max\{M_{P}^{ w_{A} w_{B}} + M_{P}^{ w_{B} b_{C}}, M_{P}^{ b_{A} w_{B}} + M_{P}^{ w_{B} b_{C}}$\}
\item $M_{R}^{ b_{B} w_{C}} = max\{M_{P}^{ w_{A} w_{B}} + M_{P}^{ b_{B} w_{C}}, M_{P}^{ b_{A} b_{B}} + M_{P}^{ w_{B} w_{C}}, M_{P}^{ w_{A} w_{B}} + M_{P}^{ b_{B} w_{C}}, M_{P}^{ b_{A} b_{B}} + M_{P}^{ w_{B} w_{C}}$\}
\item $M_{R}^{ b_{B} b_{C}} = max\{M_{P}^{ w_{A} w_{B}} + M_{P}^{ b_{B} b_{C}}, M_{P}^{ b_{A} b_{B}} + M_{P}^{ w_{B} b_{C}}, M_{P}^{ w_{A} w_{B}} + M_{P}^{ b_{B} b_{C}}, M_{P}^{ b_{A} b_{B}} + M_{P}^{ w_{B} b_{C}}$\}
\end{enumerate}
\end{itemize}

\subsection{Illustration}
\begin{figure}
\xymatrix{
&l=0 &&& l=1 && l=2 \\
a \ar@{-}[dr]_{(0,\$,\$,1)}="a1" && b \ar@{-}[dl] ^{(0,\$,\$,1)}="b1" && a \ar@{-}[d] ^{(0,1,1,\$)}="a2" && a \ar@{-}[d]^{(1,2,2,2)}  \\
& c \ar@{-}[d]^{(0,\$,\$,1)}="c1" &&& c \ar@{-}[d]^{(0,1,1,\$)}="c2" && e\\
& d \ar@{-}[d]^{(0,\$,\$,1)}="d1" &&& e\\
& e}

\caption{Top tree augmented with the information to maintain maximum-matching for the tree. Cardinality of maximum-matching for this tree is 2 as this is the maximum-number among the four constrained matchings for the cluster at top-most level of the tree.}
\label{fig:top-tree-aug}
\end{figure}
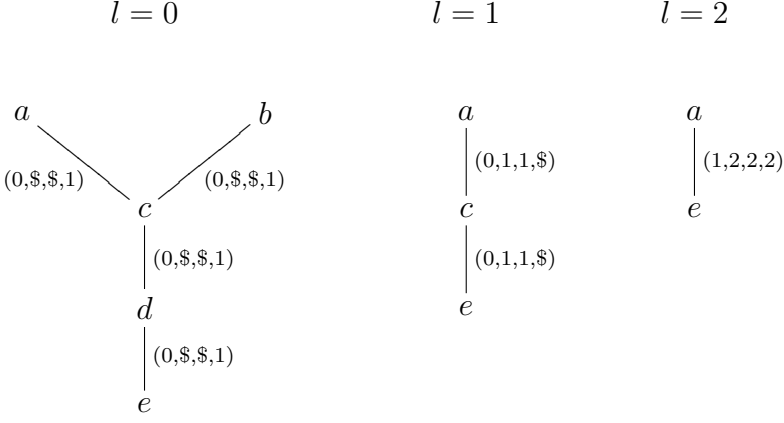

Let us take an example to illustrate the algorithm (figure \ref{fig:top-tree-aug}). We augment the top-tree of the earlier example with the information stored at each cluster. At each cluster $Q$, $X$ and $Y$ being the boundary nodes of $Q$, we store the tuple $(M_{Q}^{w_{X}w_{Y}}, M_{Q}^{w_{X}b_{Y}}, M_{Q}^{b_{X}w_{Y}}, M_{Q}^{b_{X}b_{Y}})$. For the purpose of illustration, we choose that boundary nodes as $X$ such as $X$ is situated vertically below $Y$ in the figure. For example in $(a,c)_{0}$, we choose $X=a$ and $Y=c$ as $a$ occurs above $c$ in the illustration. Further, we have represented $null$ value by \$.

\begin{figure}
\xymatrix{
&l=0 &&& l=1 && l=2 && l=3\\
a \ar@{-}[dr]_{(0,\$,\$,1)}="a1" && b \ar@{-}[dl] ^{(0,\$,\$,1)}="b1" && a \ar@{-}[d] ^{(0,1,\$,1)}="a2" && a \ar@{-}[d]^{(1,1,1,\$)}  && a\ar@{-}[d]^{(2,2,2,2)}\\
& c \ar@{-}[d]^{(0,\$,\$,1)}="c1" &&& c \ar@{-}[d]^{(0,\$,\$,1)}="c2" && d\ar@{-}[d]^{(0,\$,1,1)}&&f\\
& d \ar@{-}[dl]_{(0,\$,\$,1)}="d1" \ar@{-}[dr]^{(0,\$,\$,1)}="	d2" &&& d\ar@{-}[d]^{(0,\$,1,1)} && f\\
e&&f&&f}
\caption{Updation of top-tree along with the augmented information when the edge $(d,f)$ is added to the graph. The maximum-matching of the tree remains to be 2 as it is still the maximum number among the four constrained matching for the cluster at top-most level of the tree.}
\label{fig:top-tree-upd}
\end{figure}
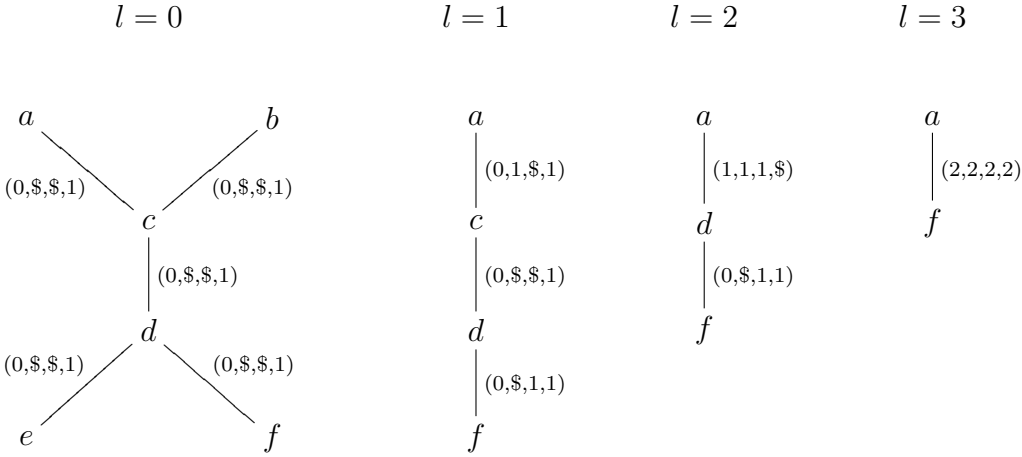

Now, on adding edge, $(d,f)$, the top-tree is updated in the same manner (figure: \ref{fig:top-tree-upd}) as it was updated earlier. In addition, we update the information stored at the clusters.

\subsection{Time Complexity and correctness}
The correctness of the algorithm can be inferred from two observations made in the section \ref{observations}. The algorithm takes $O(log(n))$ for updation since the top-tree updation takes O(log(n)) time and the algorithm, in addition, does only a constant amount of work for each updated cluster.

The query time to find the cardinality of maximum-matching is $O(1)$ since we simply need to find the maximum-number among the set of constrained matchings for the top-most cluster in the top-tree.

\subsection{Finding the matched edges}
We have so far presented an algorithm on how to maintain the cardinality of maximum-matching. However, we have not shed light on how to find whether a particular edge is matched or not. The answer to the question depends upon whether the tree in question has a unique maximum matching or not. For in case the tree in question has more than one maximum-matchings then the question has either to be modified to asking whether a particular edge is matched in one of the several maximum-matchings or whether the edge is matched in the specified maximum-matching.

We first present the algorithm for finding whether a particular edge is matched in atleast one among the several possible maximum-matchings. At each level of the top-tree, we have an ancestor of the edge in question. In this algorithm, we trace the path from the edge in question to the cluster at the top-most level in the top-tree via the ancestors of the edge. Each ancestor has four possible constrained matchings stored with it. Our algorithm chooses those constrained matchings of the ancestor which have the edge in question as matched. At the top most cluster of the top-tree, we check whether any of the chosen constrained matchings of the top-most cluster has cardinality equal to the maximum-matching. If yes, the edge is matched in atleast one of the possible maximum-matchings. If no, then the edge in question is not matched.

We start from the edge in question and choose that constrained matching, among the four available constrained matchings at level 0, which makes the edge matched. We then go one level up and check which of the four constrained matchings stored at the parent cluster can be achieved by combining the choice of matching made at level 0 with all possible valid matchings of the other child cluster of the parent. We include these achievable constrained maximum-matchings of the parent cluster in our choice set at level 1 since these constrained maximum-matchings make the edge in question matched. Iteratively, at level $i$ we have a set of choices, which can be at maximum four in number as there are only four constrained matchings, and check which of these choices can lead to the constrained matchings stored at its parent at level $i+1$ when combined through valid matchings of the other child cluster of the parent. Whichever of the constrained matchings of level $i+1$ can be achieved are made as part of our choice at level $i+1$. At the cluster at the top-most level in the top tree, in case our choice set has non-empty intersection with the constrained matching of top-most cluster which lead to maximum-matching, then the edge in question is matched otherwise not. Further, in case at any level our choice set become empty, then the edge in question is not matched.

This clearly takes $O(log(n))$ time as there are $O(log(n))$ levels in a top-tree and a constant amount of work is done at each level. Hence, to find whether an edge is matched among the potentially several maximum-matchings takes $O(log(n))$ time. 

The above algorithm can be easily tweaked to find out whether a particular edge is matched or unmatched in the particular maximum matching where the matchings of say $k$ edges are specified. Essentially, here for all the $k$ edges we go up the tree and make choice sets, which confirm to the specified matchings of these edges, at ancestors of the $k$ edges. Then we start from the edge in question and traverse up the top-tree and build the choice set where the edge in question is matched. If at any ancestor, the choice sets of the edge in question has zero intersection with the choice sets of the specified $k$ edges, the edge is unmatched. Else, if at the top-most cluster, we have a non-zero intersection, the edge is matched. This again takes $O(k*log(n))$ time.

\subsection{Tree with weighted edges}
The algorithm can be easily extended to tree with weighted edges by setting $M_{Q}^{b_{A}b_{B}}=w(A,B)$ for all edge-clusters $Q=(A,B)$ at level 0 of the top-tree where $w(A,B)$ specifies the weight of the edge $(A,B)$.

\section{Conclusion and Future Work}
This paper presented an $O(log(n))$ update time fully-dynamic algorithm for maintaining maximum-matching in a tree. 
Future work on developing dynamic algorithms for maximum-matching for general graphs in particular bipartite graphs and planar graphs.

\section{Acknowledgment}
We would like to acknowledge the effort and energy put in by Dr. Surender Baswana in this research.

\bibliography{dynamictreematching}
\end{document}